\documentclass[11pt]{amsart}

\usepackage{mathptmx}

\usepackage{amsxtra} 
\usepackage{t1enc} 
\DeclareMathAlphabet{\mathpzc}{OT1}{pzc}{m}{it} 

\newtheorem{theorem}{Theorem}
\newtheorem{lem}{Lemma}
\newtheorem{prop}{Proposition}

\newcommand{\NN}{\mathbb{N}}
\newcommand{\ZZ}{\mathbb{Z}}
\newcommand{\RR}{\mathbb{R}}
\newcommand{\TT}{\mathbb{T}}
\newcommand{\e}{\operatorname{e}}

\newcommand{\bs}[1]{\pmb{#1}}

\begin{document}

\title{Diffraction spectrum of lattice gas models above $T_c$}

\author{Michael Baake}
\author{Bernd Sing}
\address{Fakult\"{a}t f\"{u}r Mathematik,
  Universit\"{a}t Bielefeld, Postfach 10 01 31,
  \newline\hspace*{1em} 33501 Bielefeld, Germany}
\email{mbaake@mathematik.uni-bielefeld.de}
\email{sing@mathematik.uni-bielefeld.de}
\urladdr{http://www.mathematik.uni-bielefeld.de/baake/}

\begin{abstract} 
The diffraction spectra of lattice gas models on $\ZZ^d$ with
finite-range ferromagnetic two-body interactions above $T_c$ or with
certain rates of decay of the potential are
considered. We show that these diffraction spectra almost
surely exist, are $\ZZ^d$-periodic and consist of a pure point part
and an absolutely continuous part with continuous density. 
\end{abstract}

\subjclass[2000]{52C07, 42B05, 60D05, 82B20}
\keywords{Diffraction Spectra, Lattice Gases, Correlation Functions}

\maketitle

\section{Introduction}

In the following, we analyze the diffraction spectrum of translation invariant
Ising type models on $\ZZ^d$, interpreted as a lattice gas. More
specifically, we consider models with single spin space $\{-1,1\}$ and
pair potentials which can be described by a real symmetric function
$J(x)=J(-x)$ for $x\in\ZZ^d$ (so the Hamiltonian can formally be
written as $H=-\sum_{x,y\in\ZZ^d} J(x-y) \sigma^{}_x \sigma^{}_y$, 
where $\sigma^{}_x \in\{-1,+1\}$ denotes the spin at $x\in\ZZ^{d}$).

For a finite subsystem, $\TT\subset\ZZ^d$ (with periodic boundary
conditions, say), the partition function in the spin-formulation is
\begin{equation*}
Z^{}_{\beta}=\sum_{\{\bs{\sigma}\}}
\exp(\,\beta\cdot\sum_{x,y\in\TT}
J(x-y)\,\sigma^{}_x\,\sigma^{}_y\,) 
\end{equation*}
where the sum runs over all configurations
$\bs{\sigma}=\{\sigma^{}_x\mathbin|x\in\TT\}$ on $\TT$. Via
\begin{equation}\label{eq:Gibbs_measure}
\mu^{}_{\beta}(\bs{\sigma}) = \frac1{Z^{}_{\beta}}\,
\exp(\,\beta\cdot\sum_{x,y\in\TT}
J(x-y)\,\sigma^{}_x\,\sigma^{}_y\,),
\end{equation}
one defines a probability measure on the (finite) configuration space. In the
infinite volume (or thermodynamic) limit, this leads to the
corresponding \textit{Gibbs measure}, compare~\cite{Geo88,Sim93} for details.
 
The set of Gibbs measures is a non-empty simplex (see~\cite[Theorem
7.26]{Geo88}), but it need not be a singleton set. If its nature
changes as a function of $\beta$ (e.g., from a singleton to a
$1$-simplex), the system undergoes a phase transition. The point 
where this happens is characterized by  
$\beta^{}_c=1/k^{}_B T^{}_c$, the so-called \textit{inverse critical
  temperature}. In the following, we will only consider cases where
the Gibbs measure is unique, i.e., where we have a singleton
set. Since extremal Gibbs measures are ergodic \linebreak (see~\cite[Theorem 
14.15]{Geo88}), the unique Gibbs measure is ergodic, and quantities
obtained as an average over the ensemble (like the correlation
functions which we consider next) are valid almost surely for each
member of the ensemble (with respect to the Gibbs measure). 

Assuming uniqueness of the Gibbs measure, the \textit{density-density
  correlation function} $\langle n^{}_0\, n^{}_x \rangle^{}_{\beta}$
in the lattice gas interpretation of the models considered can be
deduced from $n^{}_x = \frac12 (\sigma^{}_x+1)$, where the site $x$
is occupied by a particle iff $n^{}_x=1$. This leads to the following
relationship among the correlation functions (note that we assume
  uniqueness of the Gibbs measure, so $\langle \sigma^{}_x
  \rangle^{}_{\beta}=0$): 
\begin{equation}\label{eq:dens_corr}
\langle n^{}_0\, n^{}_x \rangle^{}_{\beta} = \frac14\, (\langle \sigma^{}_0\,
\sigma^{}_x \rangle^{}_{\beta}+1). 
\end{equation}

This interpretation yields the following positive definite
\textit{autocorrelation measure} (almost surely, in the sense explained above):
\begin{equation*}
\gamma = \sum_{x\in\ZZ^d} \langle n^{}_0\, n^{}_x \rangle^{}_{\beta}\;
\delta^{}_x, 
\end{equation*}
where $\delta^{}_x$ denotes the \textit{Dirac} or \textit{point measure} at
$x$, compare~\cite[Chapter 7]{Cow95} and~\cite{B02} for details on
the autocorrelation of lattice systems. The \textit{diffraction
  spectrum} of the lattice gas model is then the Fourier transform
$\hat{\gamma}$ of its autocorrelation measure $\gamma$. The positive
measure $\hat{\gamma}$ can be uniquely decomposed as
$\hat{\gamma}=(\hat{\gamma})^{}_{\textnormal{pp}}+
(\hat{\gamma})^{}_{\textnormal{sc}}+(\hat{\gamma})^{}_{\textnormal{ac}}$   
by the Lebesgue decomposition theorem, see~\cite{RS80}. 
Here, $(\hat{\gamma})^{}_{\textnormal{pp}}$ is a \textit{pure point
  measure}, which corresponds to the Bragg part of the diffraction spectrum,
$(\hat{\gamma})^{}_{\textnormal{ac}}$ is \textit{absolutely
  continuous} and $(\hat{\gamma})^{}_{\textnormal{sc}}$
\textit{singular continuous} with respect to Lebesgue measure.

The constant part of \eqref{eq:dens_corr} results in the pure point measure
\begin{equation}\label{eq:pp}
(\hat{\gamma})^{}_{\textnormal{pp}} = \frac14\cdot \sum_{k\in\ZZ^d}
\delta^{}_k = \frac14\; \delta^{}_{\ZZ^d}
\end{equation}
in the diffraction spectrum (note that $\ZZ^d$ is self-dual), as a
consequence of \textit{Poisson's summation formula} for
distributions, see~\cite[p.\ 254]{Sch98} and~\cite{Cor89}.  
Strictly speaking, the validity of~\eqref{eq:pp} is not clear at this
stage, we have only shown that $(\hat{\gamma})^{}_{\textnormal{pp}}$
``contains'' $\frac14\; \delta^{}_{\ZZ^d}$. However, it is valid if
$\frac14 \sum_{x\in\ZZ^d} \langle \sigma^{}_0\, \sigma^{}_x
\rangle^{}_{\beta}\; \delta^{}_x$ is a null weakly almost periodic
measure, see~\cite[Chapter 11]{GLA90}. All examples we will consider
here are of this type. 

We will now show that, in addition to this pure point part,
almost surely only an absolutely continuous part is present in the diffraction
spectrum of models on $\ZZ^d$ with finite-range ``ferromagnetic''
(i.e., attractive) two-body interactions for all temperatures above
$T^{}_c$. We will also show that the same holds for models (deep) in the 
Dobrushin uniqueness regime that satisfy an additional condition on
the rate of decay of their potential. Similar observations have also been made
in~\cite{vEM92,BH00}. In view of the widespread
application of such lattice gas models to disordered phenomena in
solids, this gives a partial justification why singular continuous
spectra are usually not considered in classical
crystallography. 

\section{Fourier Series of Decaying Correlations}

Let us assume that the correlation coefficients of a model considered is
either exponentially or algebraically decaying. We first look at
exponentially decaying correlations, i.e.,
\begin{equation}\label{eq:bound}
|\langle \sigma^{}_0\, \sigma^{}_x \rangle^{}_{\beta}| \le
C\cdot\e_{}^{-\varepsilon\, \|x\|}, 
\end{equation}
where $C$, $\varepsilon$ are positive constants depending only on $\beta$ and
the model considered, and $\|\cdot\|$ denotes the Euclidean distance.

We will now deduce from the absolute convergence of
$\sum_{x\in\ZZ^d} \e_{}^{-\varepsilon\, \|x\|}$
that the exponentially decaying part of the correlation yields an absolutely
continuous part in the diffraction spectrum. From the inequality
\begin{equation*}
\frac1{\sqrt{d}}\, (|x^{}_1|+\ldots+|x^{}_d|) \le \|x\|,  
\end{equation*}
we get
\begin{equation*}
\begin{split}
\sum_{x\in\ZZ^d} \e_{}^{-\varepsilon\, \|x\|} &
      \le \sum_{x\in\ZZ^d} \e_{}^{-\frac{\varepsilon}{\sqrt{d}}\,
    (|x^{}_1|+\ldots+|x^{}_d|)}  = \left(
    \sum_{n\in\ZZ} \e_{}^{-\frac{\varepsilon}{\sqrt{d}}\,
      |n|}\right)_{}^d \\
&  = \left(\frac{\e_{}^{\varepsilon/\sqrt{d}} + 1}{\e_{}^{\varepsilon/\sqrt{d}}
    - 1}\right)_{}^{d} =
    \left(\coth\left(\frac{\varepsilon}{2\,\sqrt{d}}\right)\right)_{}^{d}.    
\end{split}
\end{equation*}

So far, we have

\begin{lem}\label{lem:abs_conv}
Let the correlation coefficients be bounded as in \eqref{eq:bound}. Then the
sum $\sum_{x\in\ZZ^d} \langle \sigma^{}_0\, \sigma^{}_x \rangle^{}_{\beta}$ 
is absolutely convergent. \qed
\end{lem}

We now consider correlations with algebraic decay of power $p>d$, i.e.,
\begin{equation}\label{eq:pow_bound}
|\langle \sigma^{}_0\, \sigma^{}_x \rangle^{}_{\beta}| \le
C\cdot\|x\|_{}^{-p} 
\end{equation}
where $C$ is positive constant. 

We observe that $|\langle \sigma^{}_0\, \sigma^{}_x
\rangle^{}_{\beta}| \le 1$ for all $x\in\ZZ^d$ and that
$\|x\|^{}_{\infty}\le\|x\|$ (where \linebreak
$\|x\|^{}_{\infty}=\max^{}_i |x^{}_i|$ denotes the maximum
norm). Therefore, we get 
\begin{equation*}
\sum_{x\in\ZZ^d} |\langle \sigma^{}_0\, \sigma^{}_x
\rangle^{}_{\beta}| \le 1+C \sum_{x\in\ZZ^d\setminus\{0\}}
\|x\|_{}^{-p} \le 1+C \sum_{x\in\ZZ^d\setminus\{0\}} \|x\|_{\infty}^{-p}.
\end{equation*}
Furthermore, we observe that there are $(2\cdot n+1)_{}^{d}-(2\cdot
n-1)_{}^{d} = \mathcal{O}(n_{}^{d-1})$ lattice points $x\in\ZZ^d$ with norm
$\|x\|^{}_{\infty}=n$ ($n\in\NN$); so there is a positive constant
$c^{}_d$ such that 
\begin{equation*}
\sum_{x\in\ZZ^d\setminus\{0\}} \|x\|_{\infty}^{-p} \le c^{}_d
\sum_{n\in\NN} n_{}^{-p+d-1} = c^{}_d\cdot\zeta(p+1-d). 
\end{equation*}
Here, $\zeta$ denotes Riemann's zeta function.

We have established the following variant of Lemma~\ref{lem:abs_conv}.

\begin{lem}
Let the correlation coefficients be bounded as in \eqref{eq:pow_bound}
with $p>d$. Then the sum $\sum_{x\in\ZZ^d} \langle \sigma^{}_0\,
\sigma^{}_x \rangle^{}_{\beta}$ is absolutely convergent. \qed
\end{lem}

With the same reasoning as in~\cite[Proposition 5]{BH00} and~\cite[Satz
3.7]{Hoe01}, we now obtain

\begin{prop}\label{prop:diff}
The diffraction spectrum of a lattice gas model on $\ZZ^d$, with correlation
coefficients bounded as in~\eqref{eq:bound}, or as
in~\eqref{eq:pow_bound} with a power $p>d$, almost surely exists, is
$\ZZ^d$-periodic and consists of the pure point part of~\eqref{eq:pp} and an
absolutely continuous part with continuous density. No singular continuous part
is present.  
\end{prop}

\begin{proof}
Lattice gas models can also be treated as weighted lattice Dirac combs (with
weight $1$ if a site is occupied by a particle and weight $0$
otherwise). So, the diffraction measure $\hat{\gamma}$ can be represented as
\begin{equation*}
\hat{\gamma} = \varrho * \delta^{}_{\ZZ^d}
\end{equation*}
with a finite positive measure $\varrho$ that is supported on a fundamental
domain of $\ZZ^d$, by an application of~\cite[Theorem 1]{B02}. This yields the
$\ZZ^d$-periodicity, which is also implied by the following more explicit
arguments. 

We have already treated the pure point part in~\eqref{eq:pp}.

Since the sum $\sum_{x\in\ZZ^d} \langle \sigma^{}_0\, \sigma^{}_x
\rangle^{}_{\beta}$ is absolutely convergent, we can view the correlation
coefficients $\langle \sigma^{}_0\, \sigma^{}_x \rangle^{}_{\beta}$ as
functions in $L^1(\ZZ^d)$. Their Fourier transforms are uniformly converging
Fourier series (by the Weierstra\ss\  M-test) and are therefore
continuous functions on $\RR^d/\ZZ^d$, see~\cite[Theorem
1.2.4(a)]{Rud62}, which are then also in $L^1(\RR^d/\ZZ^d)$. Applying
the Radon-Nikodym theorem finishes the proof. 
\end{proof}

\noindent
\textsc{Remark:} The Riemann-Lebesgue lemma~\cite[Corollary
VII.1.11]{SW71} states that if \linebreak $f\in L^1(\RR^d/\ZZ^d)$ and
if $f$ admits a Fourier series ($\langle\cdot,\cdot\rangle$ denotes the scalar
product)
\begin{equation*}
\sum_{x\in\ZZ^d} a^{}_x\,\e_{}^{2\pi i \,\langle k,x\rangle}
\end{equation*}
then $a^{}_x \to 0$ as $\|x\|\to\infty$. Therefore, an algebraic decay
of power $p\le d$ might still give an analogous result, but (possibly)
without the continuity of the Radon-Nikodym density. An example is the
diffraction of the classical Ising lattice gas on $\ZZ^2$ at the
critical point, $\beta=\beta^{}_c$, see~\cite{BH00} for details. In general,
faster decay of correlations implies that higher derivatives of the 
density exist, see~\cite[Chapter 3, \S1.1]{Kis91}. In particular, 
exponential decay implies that the Radon-Nikodym density is 
$C^{\infty}_{}$.  

\section{Uniqueness of the Gibbs Measure and Decay of Correlations}

A sufficient condition for the uniqueness of the Gibbs measure is
stated in \textit{Dobrushin's uniqueness theorem}~\cite[Theorem
8.7]{Geo88}. In the situation considered here, Dobrushin's condition
reads as follows (see~\cite[Example 8.9(2)]{Geo88}):
\begin{equation}\label{eq:dobrushin}
\beta\,\sum_{x\in\ZZ^d} \tanh|J(x)| < 1.
\end{equation}
Note that a sufficient condition for \eqref{eq:dobrushin} to hold is
$\beta\,\sum_{x\in\ZZ^d} |J(x)| \le 1$.

This observation lines up with the following well-known result for
\textit{finite range ferromagnets}: 
For such models, we associate to each site $x\in\ZZ^d$ a nonnegative
number $J(x)=J(-x)\ge 0$ (ferromagnetic interaction, i.e.,
we have an attractive lattice gas) and we suppose that
there exists an $R>0$ such that $J(x)=0$ if $\|x\|>R$
(finite range). Obviously, we have $\sum_{x\in\ZZ^d} J(x) < \infty$,
so for small $\beta$ (high temperature) we are in the Dobrushin
uniqueness regime. More precisely, one knows~\cite[Section V.5]{Ell85} that
$0<\beta^{}_c<\infty$ when $d\ge2$, and $\beta^{}_c=\infty$ for $d=1$. 

In the next step, we are interested in the decay of correlations. The
result in~\cite{Gro79} can be described succinctly by saying that the
correlations decay (at high temperatures) at the same rate as the
potential. Under Dobrushin's condition, we will now consider two
cases, exponential and algebraic decay, see~\cite[Sections 8.28 --
8.33]{Geo88}. 
\begin{lem}
\textnormal{(Exponential decay)} Suppose that, in addition to
  Dobrushin's condition, we have, for some $t>0$,
\begin{equation*}
\beta\,\sum_{x\in\ZZ^d} \e_{}^{t\,\|x\|} \, |J(x)| < \infty.
\end{equation*}
Then, the correlation coefficients are exponentially decaying, i.e.,
there are constants $0<\varepsilon,\,C<\infty$ such that
\begin{equation*}
|\langle \sigma^{}_0\, \sigma^{}_x \rangle^{}_{\beta}| \le C \cdot
 \e_{}^{-\varepsilon\,\|x\|}. 
\end{equation*}
\textnormal{(Algebraic decay)} Alternatively, suppose that for some
  $p>0$, in addition to Dobrushin's condition, we have:
\begin{equation*}
\beta\,\sum_{x\in\ZZ^d} \|x\|_{}^{p} \, |J(x)| < \infty.
\end{equation*}
Then, the correlation coefficients have an algebraic decay, i.e.,
there is a constant $0< C <\infty$ such that
\begin{equation*}
|\langle \sigma^{}_0\, \sigma^{}_x \rangle^{}_{\beta}| \le C \cdot
 \|x\|_{}^{-p}.
\end{equation*}\qed
\end{lem}
Note that for sufficiently high temperatures, the additional condition
of either case implies Dobrushin's condition. Combining this
with Proposition~\ref{prop:diff}, we get the following central result. 

\begin{theorem}
Suppose that, for a lattice gas model on $\ZZ^d$ with two-body
interaction, one of the following two conditions holds:
\begin{itemize}
\item There is a $t>0$ such that 
\begin{equation*}
\beta\,\sum_{x\in\ZZ^d} \e_{}^{t\,\|x\|} \, |J(x)| < \infty.
\end{equation*}
\item There is a $p>d$ such that
\begin{equation*}
\beta\,\sum_{x\in\ZZ^d} \|x\|_{}^{p} \, |J(x)| < \infty.
\end{equation*}
\end{itemize}
Then, for sufficiently high temperatures, the
diffraction spectrum of such a model almost surely exists, is $\ZZ^d$-periodic,
and consists of a pure point part as in~\eqref{eq:pp} and an absolutely
continuous part with continuous density. It cannot have any singular continuous
component.\qed  
\end{theorem} 

\noindent
\textsc{Remark:} Algebraic decay (of power $p>d=2$) of the
correlations occurs in certain dimer models~\cite{BH00,Hoe99} that can
be interpreted as (crystallographic) random tiling models. 

\section{Models with Finite-Range Ferromagnetic Two-Body Interactions}

For models with finite-range ferromagnetic two-body interaction, one
even gets exponential decay for all supercritical temperatures
$\beta<\beta^{}_c$ (and not only in the Dobrushin uniqueness regime
$\beta\ll\beta^{}_c$). Before stating the result, we note that \linebreak
$\langle \sigma^{}_0\, \sigma^{}_x \rangle^{}_{\beta} \ge 0$ by the
first Griffiths inequality~\cite[Section V.3]{Ell85}.

\begin{theorem}\cite[Theorem A]{CIV}\label{thm:CIV}
For a ferromagnetic model on $\ZZ^d$ with finite-range two-body interaction,
the following holds for $\beta<\beta^{}_c$ and uniformly in $\|x\| \to
\infty$\textnormal{:}
\begin{equation}\label{eq:CIV}
\langle \sigma^{}_0\, \sigma^{}_x \rangle^{}_{\beta} = \frac{\Phi^{}_{\beta}
  (\mathfrak{n}(x))}{\|x\|_{}^{(d-1)/2}}\; \e_{}^{-\|x\|\;
  \xi^{}_{\beta}(\mathfrak{n}(x))} \; (1+\mathpzc{o}(1))
\end{equation}
where $\mathfrak{n}(x)$ is the unit vector in the direction of $x$,
$\mathfrak{n}(x) = x/\|x\|$, $\Phi^{}_{\beta}$ is a strictly positive locally
analytic function on $\mathbb{S}^{d-1}$, and $\xi^{}_{\beta}$ denotes the
inverse correlation length.\qed
\end{theorem}

The inverse correlation length $\xi^{}_{\beta}$ is a positive function
by~\cite[Theorem 1]{ABF87} in connection
with~\cite[Theorems 1.3 \& 2.1]{Sim80}.

\begin{lem}\cite[Theorem 1.1]{CIV}
Under the assumption of Theorem~\ref{thm:CIV}, one has $\xi^{}_{\beta}
>0$ on $\RR^d\setminus\{0\}$ iff $\beta<\beta^{}_c$.\qed 
\end{lem}

\noindent
Note that $\xi^{}_{\beta}$ is homogeneous of degree one, i.e.,
$\xi^{}_{\beta}(\alpha\cdot x) = \alpha\cdot\xi^{}_{\beta}(x)$ for
$\alpha>0$. It also follows from the proof of~\cite[Theorem A]{CIV}
that $\xi^{}_{\beta}$ is an analytic and therefore continuous function
on $\mathbb{S}^{d-1}$, see~\cite[p.\ 309]{CIV}.

One can now modify the bound~\eqref{eq:CIV} to meet our requirements.

\begin{lem}
Under the assumption of Theorem~\ref{thm:CIV}, there are positive
constants $C$, $\varepsilon$, such that~\eqref{eq:CIV} can be bounded
as in~\eqref{eq:bound}. 
\end{lem}

\begin{proof}
Since $\xi^{}_{\beta}$ is a positive continuous function over a compact set
$\mathbb{S}^{d-1}$, we have $\inf_{x\in\mathbb{S}^{d-1}} \xi^{}_{\beta}(x) \ge
\varepsilon >0$. 

Similarly, since $\Phi^{}_{\beta}$ is locally analytic, we can also find an
appropriate $M>0$ such that
\begin{equation*}
\frac{\Phi^{}_{\beta}
  (\mathfrak{n}(x))}{\|x\|_{}^{(d-1)/2}}\; \e_{}^{-\|x\|\;
  \xi^{}_{\beta}(\mathfrak{n}(x))} \; (1+\mathpzc{o}(1)) \le
  \frac{M}{\|x\|_{}^{(d-1)/2}} \, \e_{}^{-\varepsilon\, \|x\|}. 
\end{equation*}
So far, we have an upper bound for the asymptotic behaviour of $\langle
\sigma^{}_0\, \sigma^{}_x \rangle^{}_{\beta}$. But $|\langle
\sigma^{}_0\, \sigma^{}_x \rangle^{}_{\beta}|$ is bounded by $1$, so we
can find a positive constant $C\ge M$ such that
\begin{equation*}
\langle \sigma^{}_0\, \sigma^{}_x \rangle^{}_{\beta} \le
\frac{C}{(1+\|x\|)_{}^{(d-1)/2}} \, \e_{}^{-\varepsilon\, \|x\|}
\le C\cdot\e_{}^{-\varepsilon\, \|x\|}. 
\end{equation*}
This proves the claim.
\end{proof}

It is possible to obtain the exponential decay also directly
from~\cite{ABF87} together with the Simon-Lieb
inequality~\cite{Sim80,Lie80}. However, we invoked the stronger result
of~\cite{CIV} here because it might help to analyze some finer details of
the diffraction in the future. 

Combining the last Lemma with Proposition~\ref{prop:diff}, we get our main
result. 

\begin{theorem}\label{thm:main}
For $\beta<\beta^{}_c$ \textnormal{(}$T>T^{}_c$\textnormal{)}, the
diffraction spectrum of a lattice gas model on $\ZZ^d$ with
finite-range ferromagnetic two-body interaction almost surely exists,
is $\ZZ^d$-periodic and consists of the pure point part
$(\hat{\gamma})^{}_{\textnormal{pp}} = \frac14 \; \delta^{}_{\ZZ^d}$
and an absolutely continuous part whose Radon-Nikodym density is
$C^{\infty}_{}$. No singular continuous part is present.\qed  
\end{theorem} 

One further qualitative property of the absolutely continuous
component can be extracted without making additional assumptions. In
our setting, we know inequality~\eqref{eq:bound} and also that
\begin{equation*}
\eta(x)=\langle \sigma^{}_0\, \sigma^{}_x \rangle^{}_{\beta} =\langle
\sigma^{}_0\, \sigma^{}_{-x} \rangle^{}_{\beta} =\eta(-x)
\end{equation*}
which follows from the positive definiteness of the
autocorrelation. Consequently, one has
\begin{equation*}
\left(\sum_{x\in\ZZ^d} \eta(x) \,
  \delta^{}_x\right)\sphat(k) = \sum_{x\in\ZZ^d} \eta(x) \, \cos(2\pi k x)
\end{equation*}
where the right hand side is a uniformly converging Fourier series of
a $\ZZ^d$-periodic continuous function, as consequence of
Lemma~\ref{lem:abs_conv}. In fact, in our setting of exponential decay of
$\eta(x)$, this function is $C^{\infty}$. Since $\eta(x)\ge 0$ for all
$x\in\ZZ^d$, this function has absolute maxima at $k\in\ZZ^d$ (viewed
as the dual lattice of $\ZZ^d$).

\begin{prop}
Under the assumptions of Theorem~\textnormal{\ref{thm:main}}, the absolutely
continuous component of the diffraction measure is represented by a
continuous function that assumes its maximal value at positions
$k\in\ZZ^d$.\qed
\end{prop}

This result reflects the well-known qualitative property that the
diffuse background (i.e., the continuous components) concentrates
around the Bragg peaks if the effective (stochastic) interaction is
attractive. Otherwise, the two components ``repel'' each other,
as in the dimer models, see~\cite{Cow95,Wel85} for details. 

\medskip
 
\noindent
\textsc{Remark:} All results also hold -- mutatis mutandis -- for an arbitrary
lattice $\varLambda \subset \RR^d$, since there exists a bijective linear map
$\varLambda\to \ZZ^d,\; x\mapsto Ax$ where $A\in\textnormal{GL}_d(\RR)$
(i.e., $A$ is an invertible $d\times d$-matrix with coefficients in
$\RR$). E.g., we can interpret a finite-range model on $\varLambda$ with
range $R$ as finite-range model on $\ZZ^d$ with range (bounded by)
$\|A\|^{}_2\cdot R$, where $\|\cdot\|^{}_2$ denotes the spectral norm of the
matrix $A$. The ferromagnetic two-body interaction
$\tilde{J}(x)=\tilde{J}(-x)\ge0$ on $\varLambda$ changes to
$J(y)=\tilde{J}(A^{-1}_{}y)=\tilde{J}(-A^{-1}_{}y)=J(-y)\ge0$ for
$y\in\ZZ^d$. Also, Dobrushin's uniqueness condition does not require
any specific underlying structure such as $\ZZ^d$, one can choose any
lattice, and even more general structures.  

\section{Concluding Remarks} 

The precise analysis of diffraction spectra of mixed type (e.g., with
non-trivial continuous components) has recently gained
importance. This is caused by the existence of non-periodically
ordered solids~\cite{Ste90} and the observation that structural
disorder is much more widespread than previously anticipated~\cite{Wel85}.

Our simple observation above shows that singular continuous
diffraction spectra should not be expected as the result of disorder
of (ferromagnetic) lattice gas type. It is plausible that also other
types of disorder will tend to destroy rather than create singular
continuous components. 

However, there is very good evidence (based on scaling arguments and
extensive numerical investigations~\cite{Hen99,Hoe01}) that this is
different for random tiling diffraction with non-crystallogra\-phic
symmetry in two dimensions. A rigorous proof of the latter claim would
help to better understand the role of singular continuous spectra.

\section*{Acknowledgements}

It is a pleasure to thank A.C.D.~van Enter, D.~Ioffe, Yu.~ Kondratiev
and M.~R\"{o}ckner for discussions and helpful comments. Financial
support from DFG is gratefully acknowledged.

\end{document}